\documentclass[aps,prd,reprint,superscriptaddress,showpacs,longbibliography,footnoteadded]{revtex4-1}
\usepackage[T1]{fontenc}
\usepackage{float}
\usepackage{subfigure}
\usepackage{amssymb}
\usepackage{graphicx}
\usepackage{amsmath}
\usepackage{slashed}
\usepackage{tensor}
\usepackage{hyperref}
\usepackage{epstopdf}
\usepackage{extarrows}
\usepackage{appendix}
\usepackage{color}
\usepackage{bbm}
\usepackage{caption}
\usepackage[utf8]{inputenc}
\hypersetup{
    colorlinks=true,
    linkcolor=red,
    citecolor=blue,
}

\begin{document}

\begin{flushright}
CTPU-PTC-25-39
\end{flushright}
\baselineskip=0.5cm

\title{Quasibound States and Superradiant Instability of Black Hole in Analog Gravity}
\author{Hang Liu}
\email{hangliu@sjtu.edu.cn}
\affiliation{College of Physics and Materials Science, Tianjin Normal University, Tianjin 300387, China}

\author{Hong Guo}
\email{guohong@ibs.re.kr}
\affiliation{Particle Theory and Cosmology Group, Center for Theoretical Physics of the Universe, \\
 Institute for Basic Science (IBS), Daejeon 34126, Republic of Korea}

\begin{abstract}
In this paper, we adopt continued fraction method (CFM) associated with VBK approach, which is recently developed by Vieira, Bezerra and Kokkotas, to investigate the spectrum of quasibound states (QBS) and superradiant instability of massive scalar perturbation imposed on analog rotating black hole in photon-fluid model. We analyze the effects of black hole angular velocity $\Omega_H$ and scalar field mass $\mu$ on QBS spectrum with positive and negative winding number $m=\pm1$, respectively. In addition to the fundamental frequency, we also investigate the overtones in order to disclose more distinctions of spectrum between the states of $m=\pm 1$. We show that the sign of winding number can produce notable impacts on the spectrum, particularly to the imaginary part of the spectrum. We study the superradiant instability and find that the maximum instability for a given $\Omega_H$ is not in monotonic relationship with angular velocity, which is in contrast to the case in Kerr black hole spacetime. As expected, the strength of superradiant instability can be significantly weakened by increasing the winding number. These findings imply that there exists a critical angular velocity under which the instability is strongest in parameter space, and we are supposed to find it out at $m=1$. Indeed, this max instability is found to be $\omega_{Imax}\approx 1.13374\times 10^{-5}$ related to the critical angular velocity $\Omega_H\approx1.22$.   
\end{abstract}

\date{\today}

\maketitle

\section{Introduction}

Black holes are arguably the most fascinating objects predicted by Einstein's general relativity. Since the first black hole solution was found by Karl Schwarzschild in 1915, it has been the subject of intensive research in the community of gravitational physics. It is well known that black holes are versatile objects, as they not only play a pivotal role in advancing our understanding of classical gravity~\cite{Abac2025}, but also serve as a playground where gravity interacts with quantum physics~\cite{Hawking1974,Hawking1975}, thereby offering valuable clues toward the long-sought theory of quantum gravity. Given their fundamental importance, the detection and observation of black holes are of great significance. One of the routes of detecting black holes is by gravitational waves (GWs). Since the first GWs event GW150914 was detected by LIGO Scientific Collaboration~\cite{LIGOScientific:2018mvr} and Virgo Collaboration~\cite{Abbott2016}, over a hundred black hole binary merger events ~\cite{LIGOScientific:2016lio,LIGOScientific:2016vlm,LIGOScientific:2018mvr} have been reported. Observing the shadow casted by black holes serves as another route in black holes detection, an achievement realized by Event Horizon Telescope collaboration~\cite{EventHorizonTelescope:2019ths,EventHorizonTelescope:2019ggy,EventHorizonTelescope:2022wkp}. People have proposed rich applications of GWs in the study of long-standing mysteries of our universe, such as the nature of dark energy and dark matter~\cite{Michimura2020,Nagano2019,Pierce2018,Weiner2021,Garoffolo2021,Noller2020}, also in the investigation of estimating cosmological parameter with GWs standard siren~\cite{Jin:2020hmc,Zhang2019}, testing modified theories of gravity~\cite{Nunes2020,Ma2019} and examining quantum nature of gravity~\cite{Kanno2025}, etc. These prospects are indeed promising and groundbreaking progress has already been made in black holes observations, but we are still faced with limitations with the current instruments. At present, due to the limited capability of ground-based GWs detectors, only parts of black hole properties have been tested in classical regime, including the black hole spectroscopy~\cite{Berti2025} and the recent exciting progress on the examination of Hawking's black hole area law~\cite{Abac2025}, let alone the possible probe of quantum aspects of black holes, nevertheless the current situation may be significantly improved by the future space-based GWs detectors like LISA~\cite{LISA:2017pwj}, Taiji~\cite{Hu:2017mde,Gong:2021gvw} and TianQin~\cite{Gong:2021gvw,TianQin:2015yph}. 

 The superradiance plays an important role in black hole physics and is expected to be probed by GWs detectors in future. In essence, superradiance means that the outgoing waves scattered by black holes will have larger amplitudes than the incident waves, which indicates that the energy of black holes is extracted by waves, and this is regarded as the field version of Penrose process. When the boundary conditions of bound states are imposed for the waves in black holes spacetime, i.e. QBS, the scattered and amplified waves can be trapped by an effective potential well, reflected back, and undergo repeated amplification through superradiance. This process will persist in a way like nuclear fission and finally leads to instability of the system, i.e. superradiant instability. The significance of superrdaince and relevant superradiant instability of astrophysical black holes has long been realized and has been widely and intensively studied in literature, an excellent review for this subject is given in~\cite{Brito:2015oca}. Particularly after the first groundbreaking detection of GWs which brings us contemporary gravitational wave astronomy (GWA), the superradiant instability has attracted more attentions due to its relation to the ultralight bosons which may serve as an alternative dark matter candidate~\cite{Baumann:2018vus,Hui:2016ltb}. These ultralight bosons, which belong to the regime beyond the Standard Model of particle physics~\cite{Essig:2013lka}, would be efficiently produced through the superradiant instability of rapidly rotating black holes~\cite{East:2017ovw,East:2017mrj} if they indeed exist in nature. The resulting bosons will form a classical condensate known as boson clouds around black holes, and an exciting prediction is that the boson clouds are expected to emit GWs which is likely to be detected by GWs detectors, thus opening up new ways of probing new physics beyond the Standard Model. 
 
However, observing the superradiance and superradiant instability of astrophysical black holes is still challenging even by current cutting-edge technology. Although we have made considerable progress on theoretical side, the experimental confirmation in context of astrophysical black holes is still lacking. Facing with above circumstances, an alternative strategy of studying black hole physics is provided by analog gravity which was first proposed by Unruh~\cite{Unruh:1980cg} in 1981. Rather than relying solely on the direct observation of astrophysical black holes, one may turn to experimentally accessible analog rotating black holes, which can be realized in laboratory settings. Such table-top experiments offer comparatively economical and controllable environments to probe superradiant instabilities, thereby strengthening the theoretical foundation and boosting confidence in the eventual detection of GWs from ultralight boson clouds.

The essentials of analog gravity in Unruh's seminal paper is that the propagation equation of sound waves in fluid can be formulated as a Klein-Gordon equation in curved spacetime. Consequently, sound waves experience an effective gravity. Based on this notion, we can predict that if there is a region where the velocity of the fluid is faster than local sound speed $c_s$, then the sound waves can no longer escape from this supersonic region, just as an object falling into the event horizon of black holes can never return. This concept, known as acoustic black hole in analog gravity, provides a platform to study the physics of black holes created in laboratory on earth. Over the past decades, tremendous efforts and progress have been made in this direction. Recently, Ref.~\cite{Svancara:2023yrf} reported signatures of rotating curved spacetime arising from a giant quantum vortex. The remarkable experimental results in Ref.~\cite{MunozdeNova:2018fxv,Isoard:2019buh} claimed the observation of thermal Hawking radiation and the relevant Hawking temperature of an analog black hole. In addition, recent articles regarding analog Hawking radiation can be found in~\cite{Anacleto:2019rfn,Balbinot:2019mei,eskin2021hawking}. Besides the Hawking radiation, the classical properties of analog black holes have also attracted much attentions. The quasinormal modes (QNMs) in analog black hole spacetime were theoretically discussed in~\cite{Visser:1997ux,Berti:2004ju,Cardoso:2004fi,Daghigh:2014mwa,Patrick:2020yyy}, accompanied by some recent remarkable experimental examination of QNMs in~\cite{Torres:2020tzs,Jacquet:2021scv}. Superradiance in analog systems has likewise been explored in Refs.~\cite{Basak:2002aw,Richartz:2009mi,Anacleto:2011tr,Patrick:2020baa}. On the other hand, a series of advancements in~\cite{PhysRevA.70.063615,PhysRevA.69.033602,PhysRevLett.91.240407} facilitated the development of studying analog gravity by ultracold quantum gases. To get a more comprehensive introduction of analog gravity, one can refer to~\cite{Barcelo:2005fc} for a review.

It was proposed in~\cite{PhysRevA.78.063804} that rotating analog black holes can be realized within a self-defocusing optical cavity. This analogy arises from the fact that the equations governing the nonlinear optics can be reformulated into fluid dynamics which has already been employed to conceive the notion of analog black hole since 1981~\cite{Unruh:1980cg}. In such an optical system, the interaction between a light beam and the media can be perceived as a repulsive force mediated by atoms between photons at microscopic level, leading to the formation of a “photon-fluid”. The physics of analog black hole based on the photon-fluid has been investigated from multiple perspectives, including the superradiance and the relevant superradiant instability in~\cite{PhysRevA.80.065802,Ciszak:2021xlw}, QNMs and quasiresonance of scalar perturbation~\cite{Liu:2024vde,Liu:2024wch}. Intriguingly, it has reported in~\cite{Vocke2018} that this analog black hole model has been experimentally constructed,  therefore laying the ground for studying the properties of the analog black hole model from the experimental side. In this paper, we focus on the QBS spectrum and the superradiant instability of a rotating photon-fluid analog black hole. While related aspects were addressed in~\cite{Ciszak:2021xlw}, our work presents a more comprehensive investigation aimed at uncovering additional features of the superradiant instability. To this end, we employ a precision numerical method based on the VBK approach.
  
The present work is organized as follows. In Section~\ref{sec2}, we introduce the geometry of current analog black hole and derive the equations of motion of massive scalar perturbation. In Section~\ref{sec3}, the CFM and VBK approach are introduced. In Section~\ref{sec4}, we demonstrate and analyze the properties of QBS spectrum and superradiant instability. The conclusions and discussions are given in Section~\ref{sec5}.

\section{The equations of scalar perturbations }\label{sec2}

The geometry of this analog black hole spacetime is described by the following metric~\cite{PhysRevA.78.063804,PhysRevA.80.065802,Ciszak:2021xlw} in $2+1$ dimensions
\begin{equation}
\begin{aligned}
d s^{2} =&-\left(1-\frac{r_{H}}{r}-\frac{r_{H}^{4} \Omega_{H}^{2}}{r^{2}}\right) d t^{2}+\left(1-\frac{r_{H}}{r}\right)^{-1} d r^{2}\\
&-2 r_{H}^{2} \Omega_{H} d \theta d t+r^{2} d \theta^{2},
\end{aligned}
\end{equation}
where $r_H$ stands for the radius of  event horizon, $\Omega_H$ represents the angular velocity of the black hole. For more detailed discussions on this black hole model, one can refer to Refs.~\cite{PhysRevA.78.063804,PhysRevA.80.065802,Ciszak:2021xlw}. 

The massless scalar perturbations of analog black holes  have been widely studied. While in our current model, it has been found that the effective  mass $\mu$ of scalar perturbations can be introduced by the the non-local thermo-optical nonlinearities~\cite{Marino:2019flp,Ciszak:2021xlw}, so we have the massive Klein-Gordon equation,
\begin{equation}
	\Box \rho_1-\mu^2\rho_1=\frac{1}{\sqrt{-g}}\partial_{\mu}(\sqrt{-g}g^{\mu\nu}\partial_{\mu}\rho_1)-\mu^2\rho_1=0,
\end{equation}
where $\rho_1$, which serves as the massive scalar field, is the density perturbation of optical field. To obtain the radial wave equation of the perturbation field, we perform a separation of variables for  $\rho_1$,
\begin{equation}
	\rho_1(t,r,\theta)=G(r)\Psi(r)e^{-i(\omega t -m \theta)},\label{eq1}
\end{equation}
where the integer $m$ is called the winding number, and
\begin{equation}
G(r)=\frac{1}{\sqrt{(r-r_H)\Delta(r)}}, \quad \Delta(r)=\left(1-\frac{r_H}{r}\right)^{-1}.
\end{equation}
By this separation ansatz, the massive Klein-Gordon equation can be reduced to the following radial master wave equation,
\begin{equation}
\Psi''(r)+\frac{1}{r(r-1)}\Psi'(r)+\frac{r^2}{(r-1)^2}U(\omega,r)\Psi(r)=0,
\end{equation}
in which we have set $r_H=1$ which means that $r$ is measured in units of $r_H$, and both $\omega$ and $\Omega_H$ are measured in units of $r_H^{-1}$, and
\begin{equation}
\begin{aligned}	
U(\omega,r)=&\left(\omega-\frac{m\Omega_H}{r^2}\right)^2-\left(1-\frac{1}{r}\right)\left[\frac{m^2}{r^2}+\frac{1}{2r^3}\right.\\
&\left. -\frac{1}{4r^2}\left(1-\frac{1}{r}\right)+\mu^2\right].
\end{aligned}
\end{equation}

On the other hand, if we work in the tortoise coordinate $r_\ast$ defined by $dr_\ast / dr=\Delta(r)$, the master equation can be transformed to 
\begin{align}
&\frac{d^2\Psi(r_\ast)}{dr^2_\ast}+U(\omega,r)\Psi(r_\ast)=0.
\end{align}
In present work, we focus on the quasibound states of the perturbation field, which means that the scalar waves are required to be ingoing at the event horizon and vanishing at   
infinity. This requirements serve as the boundary conditions associated to the master equation, i.e.,
\begin{equation}
\Psi \sim
\begin{cases}
   e^{-i(\omega-m \Omega_H) r_\ast}, &  r_\ast \to -\infty\quad(r\to r_H), \\
   e^{-\sqrt{\mu^2-\omega^2} r_\ast}, &  r_\ast \to +\infty\quad(r\to +\infty),
\end{cases}
\end{equation}
The spectrum of QBS are complex numbers $\omega=\omega_R+i\omega_I$, with the real part $\omega_R$ and the imaginary part $\omega_I$ representing the oscillation frequency and the damping/growing (depending on its sign) rate of the states, respectively. Obviously, we must have $\mathrm{Re}(\sqrt{\mu^2-\omega^2})>0$ to make scalar waves vanish at infinity required by boundary conditions.

\section{The Methods}\label{sec3}

In this section, we introduce two methods used to calculate the spectrum of QBS. One is the Leaver's Continued Fraction Method (CFM) which is famous for its high accuracy, and another one is the VBK approach recently developed by Vieira, Bezerra and Kokkotas~\cite{Vieira:2021ozg,Vieira:2022pxd,Vieira:2021nha,Vieira:2025ljl}. The merit of VBK approach which exploits confluent Heun functions is that it can yield exact formula of QBS frequency.

\subsection{Leaver's Continued Fraction Method}

Leaver~\cite{Leaver,PhysRevD.41.2986} first calculated the QNMs frequency by numerically solving a three-term recurrence relation, which is now well-known as CFM. This method has also been applied to calculate the QBS spectrum of massive scalar perturbation on Kerr spacetime by Dolan in~\cite{Dolan:2007mj}. One can refer to~\cite{Leaver,PhysRevD.41.2986,Konoplya:2011qq} for a detailed discussion on CFM.

We have already get the following master equation,
\begin{equation}
\Psi''(r)+\frac{1}{r(r-1)}\Psi'(r)+\frac{r^2}{(r-1)^2}U(\omega,r)\Psi(r)=0,\label{eq2}
\end{equation}
which has two regular singular points at $r=1$ and $r=0$, and one irregular singularity at $r\to\infty$. By employing  the boundary conditions of the perturbation field $\Psi(r)$, we are able to get the asymptotic solutions at the horizon $r\to r_H$ and infinity $r\to\infty$. The first boundary condition is that only the ingoing waves are allowed when $r\to r_H$, which leads to asymptotic solution
\begin{equation}
\Psi(r)\thicksim (r-1)^{-i(\omega-m\Omega_H)},\quad r\to r_H.	\label{bc_rh}
\end{equation}
The second boundary condition requires vanishing scalar waves when $r\to\infty$. In this situation, we need to be careful to get the appropriate asymptotic solution since the infinity is an irregular singularity, which indicates that we have to consider the subdominant power law behavior in addition to the dominant exponential behavior of solution in order to maintain the accuracy of CFM. To this end, we take the following ansatz of vanishing modes at infinity
\begin{equation}
\Psi(r)\thicksim	 e^{-\sqrt{\mu^2-\omega^2}r}r^{\kappa},
\end{equation}
and then substitute this formula back to Eq.~\eqref{eq2} and take a limit $r\to\infty$, such that we can get the expression of $\kappa$
\begin{equation}
\kappa=\frac{2\omega^2-\mu^2}{2\sqrt{\mu^2-\omega^2}},	
\end{equation}
which leads  to 
\begin{equation}
\Psi(r)\thicksim	 e^{-\sqrt{\mu^2-\omega^2}r}r^{\frac{2\omega^2-\mu^2}{2\sqrt{\mu^2-\omega^2}}}. \label{bc_rinf}
\end{equation}
With the asymptotic solutions, we can expand perturbation field into following Frobenius series around event horizon,
\begin{equation}
\begin{aligned}
	\Psi(r)=&e^{-\sqrt{\mu^2-\omega^2} r}r^{\frac{2\omega^2-\mu^2}{2\sqrt{\mu^2-\omega^2}}+i(\omega-m\Omega_H)}\\
	&\times (r-1)^{-i(\omega-m\Omega_H)}\sum_{n=0}^{\infty}a_n \left(\frac{r-1}{r}\right)^n.\label{expansion}
\end{aligned}
\end{equation}
By this expansion, we can get the three-term recurrence relation for the expansion coefficients, 
\begin{equation}
\begin{split}
&\alpha_0 a_1+\beta_0 a_0=0,\\
&\alpha_n a_{n+1}+\beta_n a_n+\gamma_n a_{n-1}=0, \quad n\geq1,
\end{split}
\end{equation}
where
\begin{widetext}
\begin{equation}
\begin{aligned}
\alpha_n &=4(1+n)(\mu^2-\omega^2)\left(1+n-2 i \omega+2 i m \Omega_H\right), \\
 \beta_n &=2\left[-2\mu^4+\mu^2\left(6i\omega\sqrt{\mu^2-\omega^2}-3\sqrt{\mu^2-\omega^2}-2m^2-6n\sqrt{\mu^2-\omega^2}+4i(2n+1)\omega-(2n+1)^2+10\omega^2 \right)+\right.\\
 &\left. \omega^2\left(-8i\omega\sqrt{\mu^2-\omega^2}+4\sqrt{\mu^2-\omega^2}+2m^2+8n\sqrt{\mu^2-\omega^2}-4i(2n+1)\omega+(2n+1)^2-8\omega^2\right)-\right.\\
&\left. 4im\Omega_H\left(\frac{3}{2}\mu^2\sqrt{\mu^2-\omega^2}
-2\omega^2\sqrt{\mu^2-\omega^2}+(\mu^2-\omega^2)(2n+1-2i\omega)\right)\right],\\
\gamma_n &=4 m \Omega_H \times 
 \left(i \mu^2 \sqrt{\mu^2-\omega^2}-2 i \omega^2 \sqrt{\mu^2-\omega^2}+2(\mu^2-\omega^2)(\omega+i n)\right)
 +\mu^4+ \mu^2\left(-4 i \omega \sqrt{\mu^2-\omega^2}+4 n^2+\right.\\
 &\left.4 n \sqrt{\mu^2-\omega^2}-8 i n \omega- 8 \omega^2-1\right)+\omega^2\left(8 i \omega \sqrt{\mu^2-\omega^2}-4 n^2-8 n \sqrt{\mu^2-\omega^2}+ 8 i n \omega+8 \omega^2+1\right).
\end{aligned}
\end{equation}
\end{widetext}
The ratio of successive $a_n$ is given by infinite continued fraction,
\begin{equation}
\frac{a_{n+1}}{a_n}=\frac{-\gamma_{n+1}}{\beta_{n+1}-\frac{\alpha_{n+1} \gamma_{n+2}}{\beta_{n+2}-\frac{\alpha_{n+2} \gamma_{n+3}}{\beta_{n+3}-...}}},
\end{equation}  
and for $n=0$ we have
\begin{equation}
\beta_0-\frac{\alpha_0 \gamma_1}{\beta_1-\frac{\alpha_1 \gamma_2}{\beta_2^{\prime}-\frac{\alpha_2 \gamma_3}{\beta_3-\ldots}}}=0. \label{cf}
\end{equation}
The above condition is only satisfied for bound states such that QBS spectrum can be obtained by solving Eq.~\eqref{cf} which is an equation in terms of $\omega$.

\subsection{The VBK Approach}

To implement the VBK approach, the key step is to recast the master wave equation into the form of a confluent Heun equation. To this end, we introduce a new function $R(r)$ defined by
\begin{equation}
\Psi(r)=r^{A_0}(r-1)^{A_1}e^{A_2r}R(r),\label{eq3}
\end{equation}
the exponents $A_0, A_1$ and $A_2$ are to be determined, and the new function $R(r)$ is then required to satisfy a confluent Heun equation. By substituting Eq.~\eqref{eq3} into Eq.~\eqref{eq2}, we find that this condition is fulfilled provided the exponents take the following forms
\begin{align}
A_0&=\frac{1}{2}\left(2\pm\sqrt{1-4m^2\Omega_H^2}\right),\\
A_1&=\pm i(\omega-m\Omega_H),\\
A_2&=\pm\sqrt{\mu^2-\omega^2}.
\end{align}
Taking  the boundary conditions for QBS into consideration, we find that the minus sign is the correct choice, therefore we take
\begin{align}
A_0&=\frac{1}{2}\left(2-\sqrt{1-4m^2\Omega_H^2}\right),\\
A_1&=-i(\omega-m\Omega_H),\\
A_2&=-\sqrt{\mu^2-\omega^2}.
\end{align}
The equation for $R(r)$ is now given by
\begin{equation}
\begin{aligned}
	R''(r)+&\left(2A2+\frac{2A_0-1}{x}+\frac{1+2A1}{x-1}\right)R'(x)\\
	&+\left(\frac{A_3}{x}+\frac{A_4}{x-1}\right)R(r)=0,\label{eq4}
\end{aligned}	
\end{equation}
where
\begin{equation}
\begin{aligned}
A_3=&\frac{1}{2}+A_1-A_2+A_0\left(2A_2-2A_1-1\right)+m^2\left(1+2\Omega_H^2\right)\\
A_4=&-\frac{1}{2}+A_0+A_2+A_1\left(2A_0+2A_2-1\right)-\mu^2+2\omega^2\\
&-m^2\left(1+2\Omega_H^2\right).
\end{aligned}	
\end{equation}
Comparing this equation with confluent Heun equation which is given by
\begin{equation}
y''(x)+\left(\alpha+\frac{1+\beta}{x}+\frac{1+\gamma}{x-1}\right)y'(x)+	\left(\frac{\xi}{x}+\frac{\zeta}{x-1}\right)y(x)=0,
\end{equation}
we find that equation satisfied by $R(r)$ has exactly the same form as confluent Heun equation if  we make following identifications 
\begin{equation}
	\alpha=2A_2, \beta=2(A_0-2),\gamma=2A_1,
\end{equation}
and
\begin{equation}
	\xi=A_3,\qquad \zeta=A_4.
\end{equation}
Now we can recast Eq.~\eqref{eq4} as 
\begin{equation}
R''(r)+\left(\alpha+\frac{1+\beta}{r}+\frac{1+\gamma}{r-1}\right)R'(r)+	\left(\frac{\xi}{r}+\frac{\zeta}{r-1}\right)R(r)=0.
\end{equation}
The solution to this equation is the confluent Heun functions,
\begin{equation}
R(r)=\mathrm{HeunC}(\alpha,\beta,\gamma,\delta,\eta;r)	
\end{equation}
in which the parameters are related by
\begin{align}
\xi=&\frac{1}{2}(\alpha-\beta-\gamma+\alpha\beta-\beta\gamma)-\eta,\\
\zeta=&\frac{1}{2}(\alpha+\beta+\gamma+\alpha\gamma+\beta\gamma)+\delta+\eta.
\end{align}

According to VBK approach, the spectrum of QBS satisfies following condition
\begin{equation}
\frac{\delta}{\alpha}+\frac{\beta+\gamma+2}{2}+n=0,	
\end{equation}
where $n=0,1,2,...$ denotes the overtone number. This implies that QBS spectrum can be obtained by solving the following equation of $\omega$
\begin{equation}
1+n+\frac{\mu^2-2\omega^2}{2\sqrt{\mu^2-\omega^2}}-\frac{1}{2}\sqrt{1-4m^2\Omega_H^2}-i(\omega-m\Omega_H)=0.	\label{eq5}
\end{equation}
For a more comprehensive account of the VBK approach and its applications, we refer the reader to Refs.~\cite{Vieira:2021ozg,Vieira:2022pxd,Vieira:2021nha,Vieira:2025ljl} and references therein. 

\subsection{The comparison between CFM and VBK approach}

In this subsection, we make comparisons between CFM and VBK approach in order to find out to what extent we can trust the VBK approach. 
The most interesting discovery is that when $m=n=0$, we find that the spectrum of QBS obtained by VBK approach and CFM are almost identical, as we have demonstrated in Table~\ref{tab1} which shows a rather high consistency of the results from the two methods.  This fact may serve as the evidence of the validity of the VBK and CFM.
\begin{table}[!htbp]
\centering
{
        \begin{tabular}{cccc}
    \hline\hline
    $\mu$ &Method&  $\mathrm{Re}(\omega)$ & $\mathrm{Im}(\omega)$       \\
    \hline
    $0.1$ &VBK&     $ 0.09956744880286346$ & $-0.0001728254896776613$        \\
        \cline{2-4}
        &CFM&     $ 0.09956744880286351 $  & $-0.0001728254896776192$       \\
        \hline
    $0.3$ &VBK&   $ 0.2943998557383356 $   & $-0.006947016076715476$    \\
        \cline{2-4}
        &CFM&     $ 0.2943998557383371 $   & $-0.006947016076716305$    \\
        \hline
    $0.5$ &VBK&   $ 0.488523348128175 $   & $-0.027094571493907236$   \\
        \cline{2-4}
        &CFM&     $ 0.488523348128175 $   & $-0.027094571493906414$    \\
        \hline
    $0.8$ &VBK&   $ 0.7869080056139744 $  & $-0.07895373545621676$      \\
        \cline{2-4}
        &CFM&     $ 0.7869080056139744 $  & $-0.07895373545621726$      \\
        \hline
    $1.2$ &VBK&   $ 1.2026554991113303 $ & $-0.17857615086839682$       \\
        \cline{2-4}
        &CFM&     $ 1.2026554991113283 $  & $-0.17857615086839607$      \\
        \hline\hline
\end{tabular}
}
\caption{The fundamental QBS frequencies obtained by VBK approach and CFM at $m=n=0$ and $\Omega_H=1$ for different mass values $\mu$.}\label{tab1}
\end{table}

However, for the case $m\neq0$, a noticeable discrepancy of the QBS spectrum from the two methods takes place. We list the QBS spectrum for $m=1$ with different overtone numbers in Table~\ref{tab2}, which clearly shows that VBK approach and CFM gives discrepant results, particularly for the imaginary parts $\omega_I$ of the frequency. This large mismatch in  $\omega_I$ may imply the failure of VBK, but the real parts $\omega_R$ of the spectrum from the two methods  are consistent with each other within acceptable differences. 

Despite in the $m\neq0$ case, VBK approach only works well in the calculation of $\omega_R$ as we have shown, it is still very useful in the sense that $\omega_R$ given by VBK approach can be used as initial data in the  finding of the QBS spectrum by CFM whose performance  is sensitive to the initial guessing frequency inputted by hand in our numerical code. So, with the assistance of VBK approach, we can get a much improvement of efficiency in calculating QBS spectrum by CFM. 

\begin{table}[!htbp]
\centering
{\setlength{\tabcolsep}{12pt}
        \begin{tabular}{cccc}
    \hline\hline
    $n$ &Method&  $\mathrm{Re}(\omega)$ & $\mathrm{Im}(\omega)$       \\
    \hline
    $0$ &VBK&     $ 1.15842 $ & $ -0.0833961 $        \\
        \cline{2-4}
        &CFM&     $ 1.14923 $  & $ -0.0001546 $       \\
        \hline
    $1$ &VBK&     $ 1.1741 $   & $ -0.0263357 $    \\
        \cline{2-4}
        &CFM&     $ 1.17503 $   & $ -0.0001154$    \\
        \hline
    $2$ &VBK&     $ 1.18399 $   & $ -0.0110409 $   \\
        \cline{2-4}
        &CFM&     $ 1.18562 $   & $ -0.0000672322 $    \\
        \hline
    $3$ &VBK&     $ 1.1895 $  & $ -0.00548354 $      \\
        \cline{2-4}
        &CFM&     $ 1.19076 $  & $ -0.0000395744 $      \\
        \hline
    $4$ &VBK&    $ 1.19271 $ & $ -0.00306504 $       \\
        \cline{2-4}
        &CFM&     $ 1.1936 $  & $ -0.0000245378 $      \\
        \hline
    $5$ &VBK&    $ 1.19469 $ & $ -0.00186823 $       \\
        \cline{2-4}
        &CFM&     $ 1.19531 $  & $ -0.0000160408 $      \\
        \hline\hline
\end{tabular}
}
\caption{The fundamental and overtones of QBS frequencies for $m=1$ obtained by VBK approach  and CFM  at $\mu=1.2$ and $\Omega_H=1$.}\label{tab2}
\end{table}

\section{The spectrum of QBS and superradiant instability}\label{sec4}

\subsection{The QBS Frequencies}

\begin{figure}
\centering
\includegraphics[height=2in,width=3in]{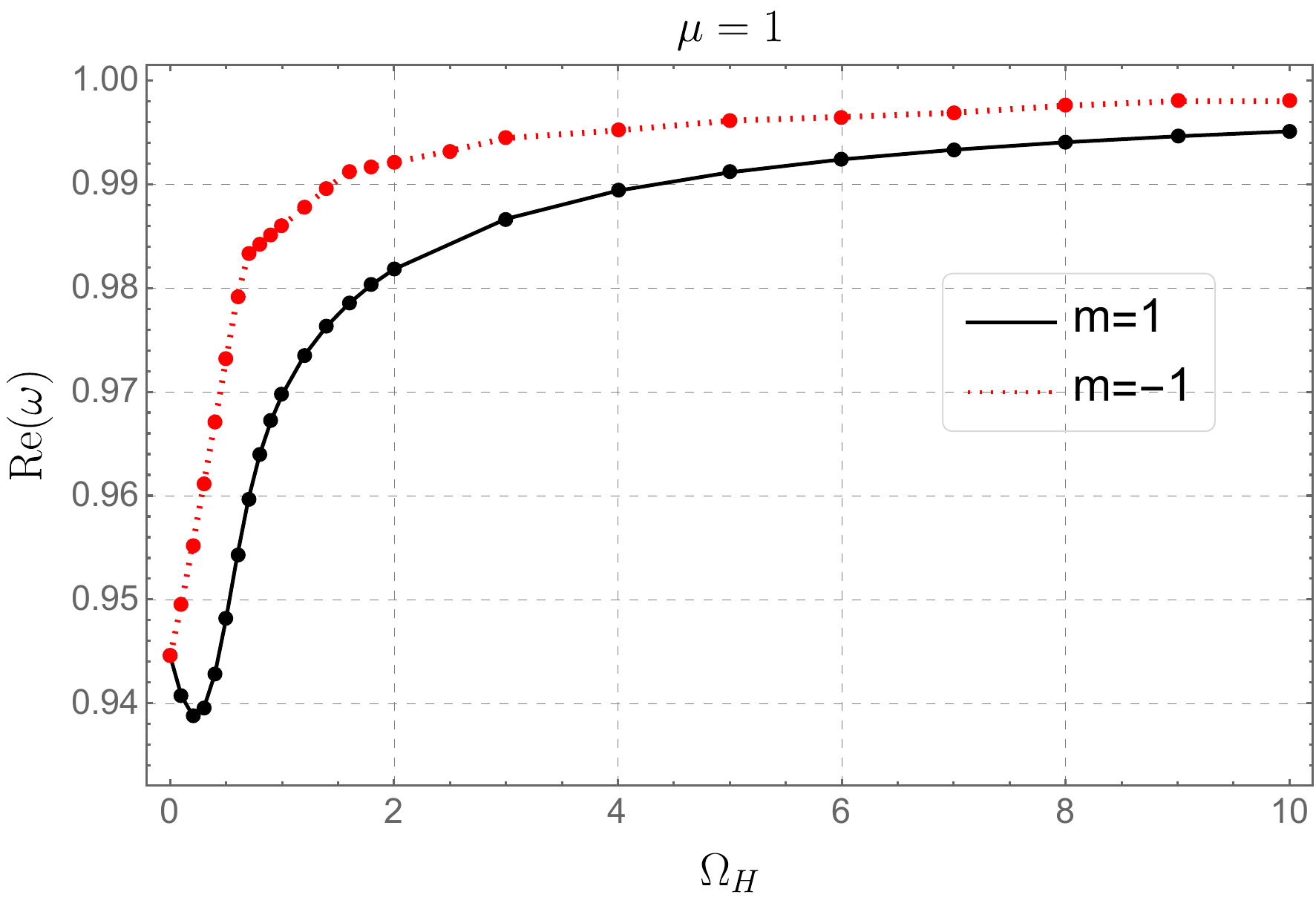}
\includegraphics[height=2in,width=3in]{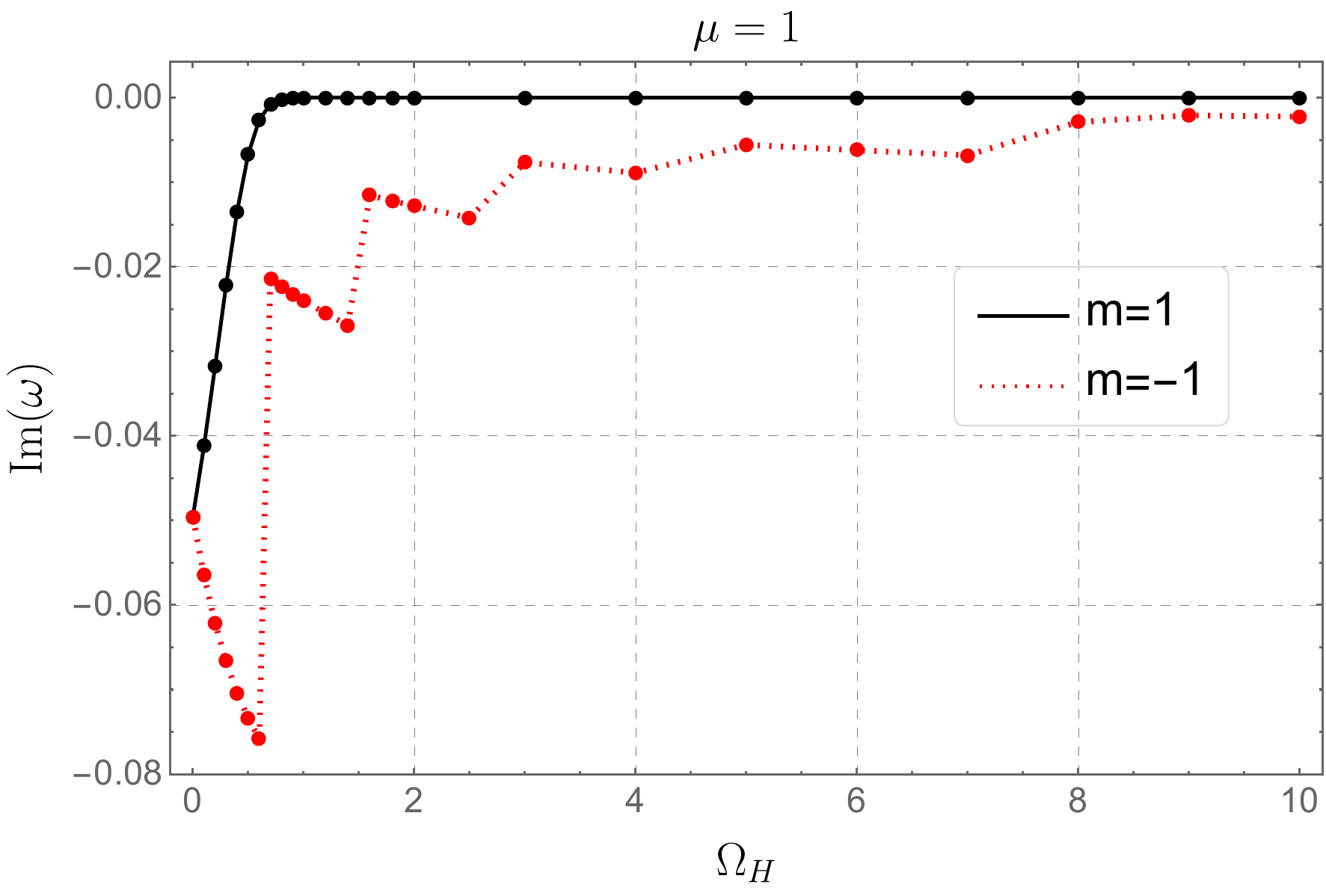}
\caption{The dependence of fundamental QBS spectrum frequencies on angular velocity $\Omega_H$ at fixed scalar mass $\mu=1$ for winding number $m=\pm1$.}\label{fig1}
\end{figure}

In this subsection, we discuss the properties of QBS spectrum. In Fig.~\ref{fig1}, we show the behavior of fundamental QBS frequency under the change of angular velocity for a fixed scalar mass $\mu=1$, and the real and imaginary parts of the spectrum are separately illustrated in the upper and lower plot, respectively. To reflect the distinct features of the spectrum related to positive and negative winding number $m$, we also include a comparison of spectrum between $m=1$ (co-rotating states) and $m=-1$ (counter-rotating states). We can see that the QBS spectrum for positive and minus winding number has distinctively different behaviors. The $\omega_R$ with $m=-1$ has generally larger values than $\omega_R$ for $m=1$, with the exception that frequencies coincide when $\Omega_H=0$ as the consequence of ``azimuthal'' degeneracy (winding number $m$ acts as azimuthal number of states in Kerr spacetime) which is broken by the introduction of black hole rotation, as we have shown in the plots. When increasing the $\Omega_H$ from zero, $\omega_R$ for co-rotating states manifests a quick drop and then monotonously increase, while the counter-rotating states just monotonously become greater. On the other hand, when we keep increasing angular velocity,  the $\omega_R$ of both co-rotating and counter-rotating states seem to get more and more close to $\mu=1$ but never exceed it, i.e. we have restriction $\omega_R<\mu$. 

For $\omega_I$ demonstrated in the lower plot of Fig.~\ref{fig1}, which stands for the growing or damping rate of the states and has apparently different behaviors from the $\omega_R$. We can observe that the co-rotating states have larger $\omega_I$ than counter-rotating states at nonzero $\Omega_H$ region, and both states have identical $\omega_I$ value when $\Omega_H=0$ due to the azimuthal degeneracy again. For the negative $\omega_I$, a higher value (smaller magnitude) implies a slower damping rate of the states. When we increase $\Omega_H$, the co-rotating states keep their $\omega_I$ monotonously growing and finally approach  zero (even exceed zero and become positive). While the behavior of $\omega_I$ for counter-rotating states is a bit confusing, as one can see several sharp drop and up of the $\omega_I$ values which still tend to get close to zero as the angular velocity grows. Actually, for the QBS with positive $m$, when black holes rotate fast enough to reach  $m\Omega_H\gtrsim\mu$, the $\omega_I$ usually become positive indicating the occurrence of  superradiant instability (we leave this topic to next subsection), and this phenomenon will never happen for counter-rotating states (negative $m$) if we restrict $\omega_R>0$. At this stage, we can conclude that the co-rotating states oscillate with a lower frequency than counter-rotating states which fade away faster than co-rotating states. 

\begin{figure}
\centering
\includegraphics[height=2in,width=3in]{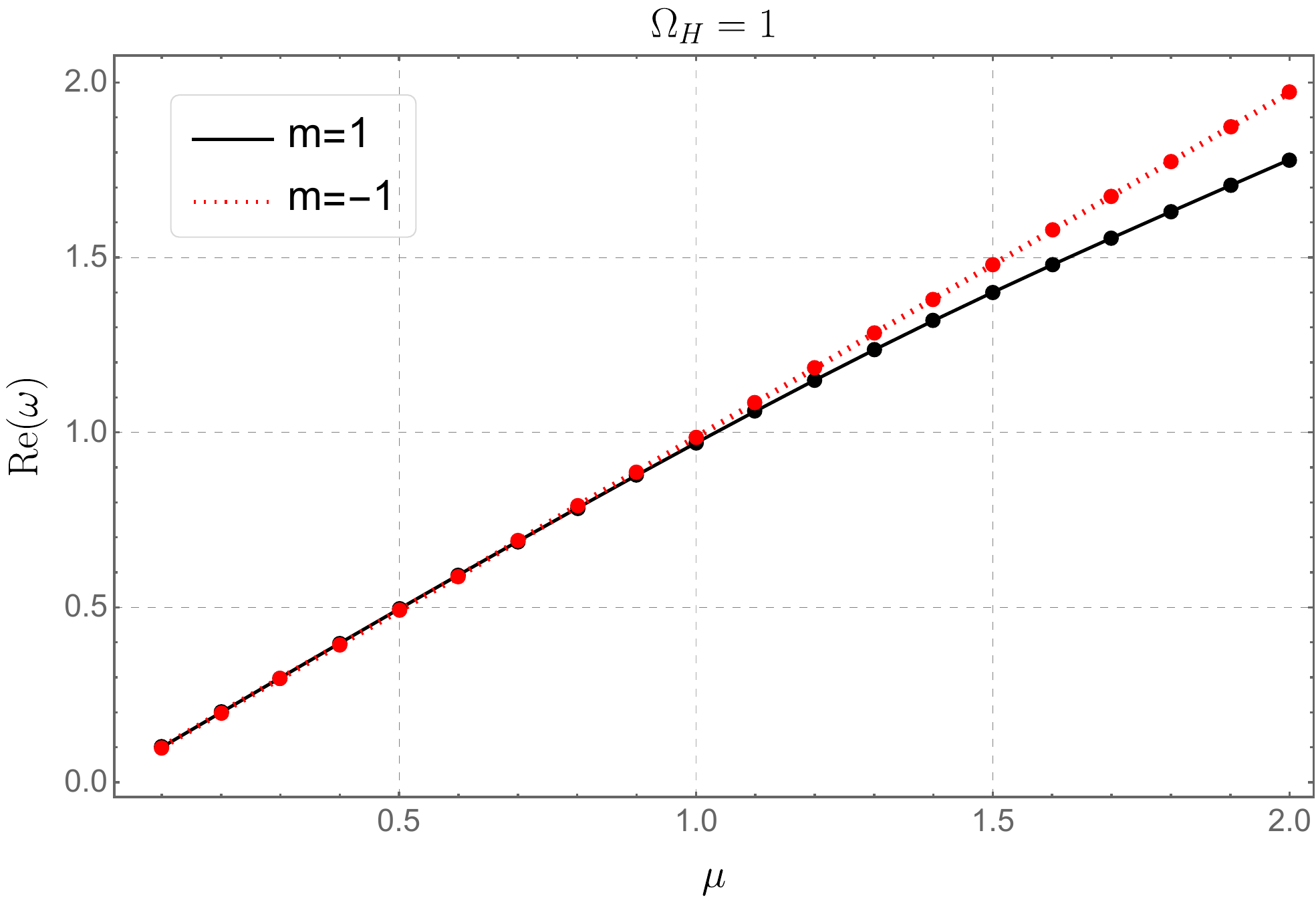}
\includegraphics[height=2in,width=3in]{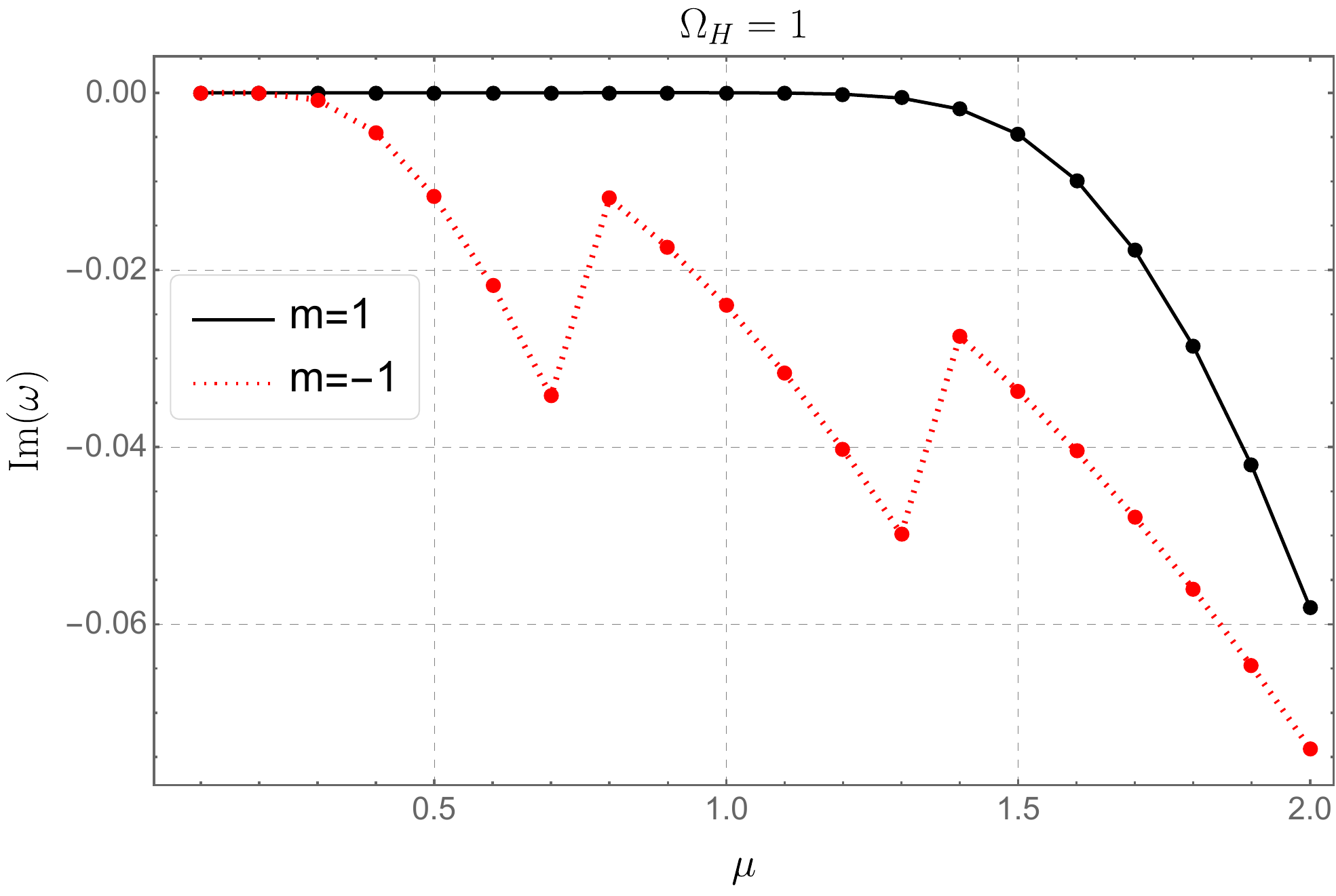}
\caption{The dependence of  QBS spectrum on scalar mass $\mu$ at fixed angular velocity $\Omega_H=1$ for winding number $m=\pm1$.}\label{fig2}
\end{figure}

In Fig.~\ref{fig2}, we show the fundamental spectrum curves as a function of scalar mass $\mu$ by fixing $\Omega_H=1$. In this scenario, we find that the $\omega_R$ of both co-rotating and counter-rotating states monotonously increase with the scalar mass in a seemingly linear way, and we find that the relation $\omega_R<\mu$ found in Fig.~\ref{fig1} still holds here. The differences of $\omega_R$ induced by different winding number $m$ are enhanced by the grow of scalar mass. When it comes to $\omega_I$, we can observe completely different  behaviors compared with $\omega_R$. For the co-rotating states, their frequencies monotonously decrease with the increase of scalar mass. However, the counter-rotating states  manifest some sharp frequency drops and increases again as we have seen in Fig.~\ref{fig1}. The counter-rotating states seem to be more sensitive to the affects of scalar mass than co-rotating states, this is due to the fact that co-rotating states undergo superradiant instability related to a tiny positive $\omega_I$ in the mass region $\mu\lesssim 1$ in which the $\omega_I$ appear to be zero in the plot. Finally, this figure leads us to the conclusion that the co-rotating states with a larger mass will oscillate more rapidly and decay faster (outside of the superradiant region), the same result for $\omega_R$ can also be concluded for counter-rotating states whose damping rates exhibit non-monotonic relationship with scalar mass. 

The above discussions are concentrated on the fundamental QBS frequencies, we now turn to the overtones. When using Eq.~\eqref{eq5}, we find that restriction $\omega_R<\mu$ holds for positive winding numbers, while it will be violated for negative winding numbers as it is possible to get $\omega_R>\mu$. To make the argument more solid, we list overtones of QBS spectrum obtained by CFM in Table~\ref{tab4} for $m=\pm1$ as an instance. In the case of $m=1$, the higher overtones have larger $\omega_R$ and $\omega_I$, and all the $\omega_R$ are limited to $\omega_R<\mu$. For $m=-1$, the same behavior of frequencies can be found from $n=0$ to $n=13$. However, an exception shows up for overtone $n=14$ which has $\omega_R>\mu$ and a substantially decreased $\omega_I$. 

\begin{table}[!htbp]
\centering
{\setlength{\tabcolsep}{6pt}
        \begin{tabular}{cccc}
    \hline\hline
    $n$ &  $\omega\,(m=1)$ & $\omega\,(m=-1)$       \\
    \hline
    $0$ & $ 1.14923-0.0001546i $  & $ 1.18324 -0.0402436i $       \\
        \hline
    $1$ &    $ 1.17503-0.0001154i $   & $1.18571 -0.0171512i $    \\
        \hline
    $2$ &  $ 1.18562-0.0000672322i $   & $ 1.18971 -0.00850025i $    \\
        \hline
    $3$ &    $ 1.19076-0.0000395744i $  & $ 1.19258 -0.00471002i $      \\
        \hline
    $4$ &  $ 1.1936-0.0000245378i $  & $ 1.19451 -0.00284132i $      \\
        \hline
    $5$ &  $ 1.19531-0.0000160408i $  & $ 1.19581 -0.0018307i $      \\
         \hline
    $6$ &  $ 1.19643 -0.0000109801i $  & $ 1.19672 -0.0012424i $      \\
       \hline
    $7$ &  $1.19719 -7.81215\times10^{-6}i  $  & $1.19737 -0.00087903i  $      \\
   \hline
    $8$ &  $1.19773 -5.74067\times10^{-6}i $  & $ 1.19785 -0.00064345i $      \\
   \hline
    $9$ &  $1.19813 -4.33455\times10^{-6}i  $  & $ 1.19821 -0.000484468i $      \\
   \hline
    $10$ &  $ 1.19843 -3.34903\times10^{-6}i$  & $ 1.19849 -0.000373508i $      \\
   \hline
    $ 11$ &  $ 1.19867 -2.63902\times10^{-6}i $  & $1.19871 -0.000293822i  $      \\
     \hline
    $12 $ &  $ 1.19886 -2.1152\times10^{-6}i $  & $ 1.19888 -0.000235176i $      \\
     \hline
    $ 13$ &  $ 1.19913 -1.41798\times10^{-6}i $  & $ 1.19924 -0.000131045i $      \\
     \hline
    $14 $ &  $ 1.19923 -1.18206\times10^{-6}i $  & $ 1.22288 -0.118884i $      \\
     
       \hline\hline
\end{tabular}
}
\caption{The fundamental and overtones of QBS frequencies for $m=\pm1$ at $\mu=1.2$ and $\Omega_H=1$.}\label{tab4}
\end{table}


\subsection{Superradiant Instability}

In this subsection, we discuss an interesting phenomenon called superradiant instability which is related to the QBS spectrum with $\omega_I>0$ suggesting an exponentially growth of states. To make superradiant instability occur, some conditions are required. The Fig.~\ref{fig1} has shown that for a fixed scalar mass, the supperradiant instability happens when the black hole rotates fast enough. On the other hand, as shown in Fig.~\ref{fig2}, when black hole rotating speed is fixed, the superradiant instability takes place when scalar mass is limited in a range $0<\mu\lesssim1.033$, i.e. the scalar field can not be too heavy.

\begin{figure}[htbp]
\centering
\includegraphics[height=2in,width=3in]{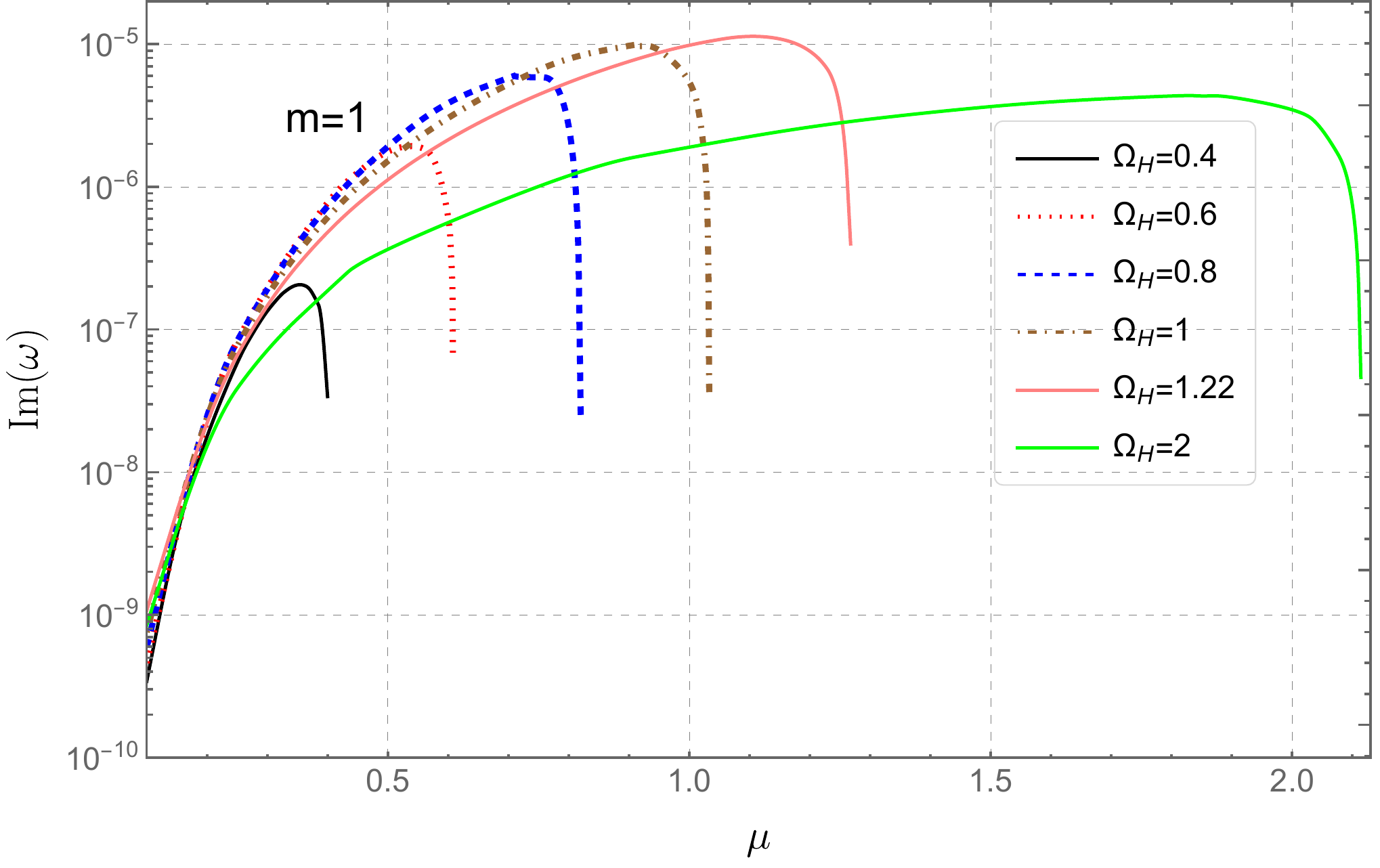}
\includegraphics[height=2in,width=3in]{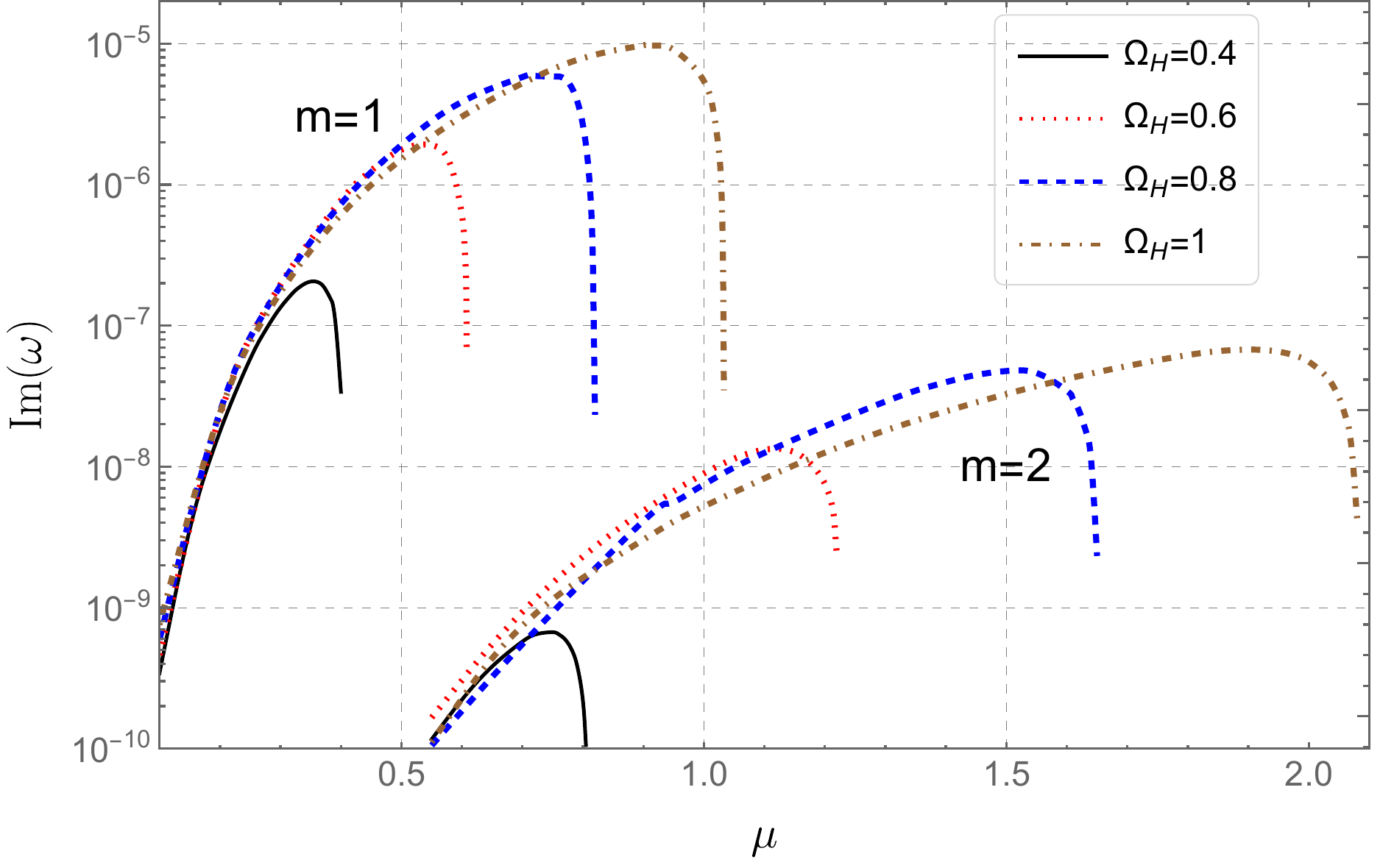}
\caption{The comparison of QBS spectrum with positive $\omega_I$ between different angular velocity $\Omega_H$ and winding number $m$.}\label{fig3}
\end{figure}

In Fig.~\ref{fig3}, we show the imaginary part $\omega_I$ of the QBS spectrum and make comparisons between different angular velocity and winding number. In the upper plot we present the $\omega_I$ for $m=1$ to reveal the effects of angular velocity on instability, and the impacts of winding number are demonstrated in the lower plot. The positive $\omega_I$ presented in the plots means that the states are experiencing superradiant instability, and the states  with a greater value of $\omega_I$ will grow faster. From the upper plot, we can see that with the increase of scalar mass, the $\omega_I$ for all angular velocity will reach  its corresponding maximum at $\mu\lesssim m\Omega_H$. In the Kerr black hole spacetime, it has been found that faster rotation creates greater instability~\cite{Dolan:2007mj}, i.e. larger maximum $\omega_I$. However, in this analog rotating black hole model, we find that the dependence of maximum growth rate on black hole rotation is not monotonic. As shown in the upper plot, the maximum $\omega_I$ of each $\Omega_H$ improves with the angular velocity from $\Omega_H=0.4$ to $\Omega_H=1.22$, but a smaller maximum growth rate is found for a even larger $\Omega_H=2$. On the other hand, the comparison between $m=1$ and $m=2$ in the lower plots shows that the instability is significantly suppressed by larger winding number. These facts may suggest a critical $\Omega_H$ at which the black hole suffers greatest instability when $m=1$. Actually, we indeed  find that the critical angular velocity is around $\Omega_H=1.22$ (pink solid curve in the figure) under which the black hole is most unstable in the sense that the growth rate of states take its max value $\omega_{Imax}\approx 1.13374\times 10^{-5}$ at $\mu\approx 1.111$ and $m=1$ in the parameter space.  

\begin{table*}[!htbp]
\caption*{$m=1$}
\centering
{\setlength{\tabcolsep}{6pt}
        \begin{tabular}{|ccc|ccc|ccc|}
    \hline
   \multicolumn{3}{|c|}{$\Omega_H=0.5$}& \multicolumn{3}{|c|}{$\Omega_H=1$}& \multicolumn{3}{|c|}{$\Omega_H=2$}\\
    \hline
    $\mu$ &  $\omega_R$ & $\omega_I(\times10^{-6})$  & $\mu$ &  $\omega_R$ & $\omega_I(\times10^{-6})$  & $\mu$ &  $\omega_R$ & $\omega_I(\times10^{-6})$  \\
    \hline

$0.1$ & $0.0999476$ & $0.000438057$ & $0.2$ &     $ 0.199641 $ & $ 0.0239801$   &   $0.5$ &  $ 0.496742 $ &  $ 0.36455 $  \\
        
$0.2$ & $0.199599$ & $0.0210394$ & $0.4$ &     $ 0.397523 $ & $ 0.609402 $   &   $0.8$ &  $ 0.78948 $ &  $ 1.19675 $\\
            
$0.3$ &$0.298689$ & $0.18116$ & $0.6$ &    $ 0.592501 $ & $ 3.03637 $     &   $1$ &  $ 0.981811 $ &  $ 1.89649 $\\
      
$0.4$ &$0.396961$ &$0.627685$  & $0.8$ &    $ 0.783601 $ & $ 7.88993 $     &   $1.2 $ &  $ 1.17161 $ &  $ 2.6243 $\\
    
$0.42$&$0.416493$ &$0.716723$  & $0.9$ &    $ 0.877366 $ & $ 9.74243 $     &   $1.4 $ &  $ 1.35865 $ &  $ 3.32229 $\\
    
$0.43$&$0.426242$ & $0.748855$ & $0.91$ &    $ 0.886668 $ & $ 9.74765 $    &   $1.6 $ &  $ 1.54265 $ &  $ 3.93754 $\\

$0.44$&$0.43598$ & $0.767328$ & $0.92$ &    $ 0.895956 $ & $ 9.69221 $    &   $1.8 $ &  $ 1.72332 $ &  $ 4.33274 $\\
     
$0.445$&$0.440844$&$0.769818$  & $0.95$ &    $ 0.923732 $ & $ 9.04906 $    &   $1.85 $ &  $ 1.76792 $ &  $ 4.33308 $\\

$0.446$&$0.441816$ &$0.76968$  & $1$ &       $ 0.969721 $ & $ 5.42582 $    &   $1.855 $ &  $1.77237 $ &  $ 4.32875 $\\
    
$0.48$ &$0.474804$ & $0.576335$ & $1.033$ &   $ 0.999853 $ & $ 0.0347234 $  &   $1.9 $ &  $ 1.81229 $ &  $ 4.23965 $\\
      
$0.5$ &$0.494137$ &$0.183665$ & $1.03315$ & $ 0.99999 $ & $ 0.00239205 $  &   $2 $ &  $ 1.90026 $ &  $ 3.4692 $\\
      
$0.506$&$0.499925$ & $0.00256938$  & $1.03316$ & $ 0.999999 $ & $ 0.000233717 $&   $2.05 $ &  $1.94386$ &  $ 2.49368 $\\
       
$0.506077$ &$0.49999954$ &$0.0000158926$  & $1.033162$ &$ 1.0000008 $ & $-0.000197994$&   $2.114855 $ &  $ 1.99999 $ &  $ 0.0000950621 $\\
    
 $0.506078$&$0.500001$ &$-0.0000173096$  &  $1.1$ &  $ 1.06044 $ & $ -25.6012$ & $2.114857 $ &  $ 2.0000002 $ &  $ -0.0000132998 $\\
       
 $0.6$&$0.589876$ &$-12.7445$  &  $1.2$ &  $ 1.14923 $ & $ -154.642$ & $2.2 $ &  $ 2.07296 $ &  $ -7.60993 $ \\
     
 $0.8$&$0.774984$ &$-599.731$  &  $1.5$ &  $ 1.4003 $ & $ -4681.27$ & $ 2.5$ &  $ 2.32214 $ &  $ -274.43 $ \\

\hline\hline
\end{tabular}
}
\caption{The fundamental spectrum of QBS with winding number $m=1$. The frequencies are grouped into three groups by value of angular velocity.}\label{tab3}
\end{table*}

In order to have a more clear picture of the superradiant instability, we list the fundamental QBS spectrum with high precision for various black hole parameters in Table~\ref{tab3}, by which we see that all the real components of QBS frequencies are constrained by $\omega_R<\mu$. When superradiant instability happens, we notice that further conditions $\omega_R<\mu<m\Omega_H+\Omega'$ and $\omega_R<m\Omega_H$ are required, where $\Omega'$ is some small positive value compared to $m\Omega_H$. For each angular velocity, the corresponding $\omega_I$ will reach its maximum at $\mu\sim m\Omega_H$. Once $\omega_R>m\Omega_H$, positive $\omega_I$ disappears and negative $\omega_I$ will show up thereby the states will decay over time and instability is absent. For $\omega_R\sim m\Omega_H$, the absolute value of $\omega_I$ will become extremely small, thus we can predict that $\omega_I\rightarrow0$ when $\omega_R\rightarrow m\Omega_H$. Obviously, the QBS with vanishing $\omega_I$ will neither grow nor decay, they just oscillate and form stationary scalar clouds surrounding the black hole. 
 
\section{Conclusions and Discussions}\label{sec5}

In this paper, we have investigated the properties of QBS spectrum and superradiant instability of massive scalar perturbation in an analog rotating black hole spacetime from the photon-fluid model. The complex QBS frequency is calculated by CFM associated with VBK approach. The characteristics of the spectrum are explored by analyzing the impacts of black hole angular velocity and scalar mass on the frequencies of QBS with positive and negative winding number ($m\pm1$). Note that the angular velocity is not limited as in the conventional rotating black holes case, e.g. Kerr black holes whose angular velocity is limited by weak cosmic censorship. This unique property of analog rotating black hole allows us to explore the QBS spectrum at larger angular velocity. 

We first fix the scalar mass $\mu=1$ and change the $\Omega_H$ in Fig.~\ref{fig1} to reveal the impacts of angular velocity on the spectrum. When $\Omega_H=0$, we find that the co-rotating states ($m=1$) and counter-rotating states ($m=-1$) have exact the same QBS frequency due to the azimuthal degeneracy which is broken by nonzero $\Omega_H$. It is found that $\omega_R$ of counter-rotating states monotonously increase with the angular velocity, while for co-rotating states, the $\omega_R$ will decrease at the beginning and then start to grow with $\Omega_H$. In the whole $\Omega_H>0$ region, counter-rotating states have larger $\omega_R$ than co-rotating states. For the imaginary part of frequency, the $\omega_I$ of co-rotating states continuously increase until it appears to approach zero. For counter-rotating states, the $\omega_I$ behaves like a step function which is kind of strange compared to the co-rotating states, nevertheless it still has a tendency to approach zero at large $\Omega_H$ as in co-rotating states. On the other hand, co-rotating states have larger $\omega_I$ than counter-rotating states. As a result, we can see that the co-rotating states oscillate with a lower frequency than counter-rotating states which fade away faster than co-rotating states, and QBS in this analog black hole spacetime with greater angular velocity will oscillate more fast with a tendency to decay slower. 

The effects of scalar mass on QBS spectrum has been illustrated in Fig.~\ref{fig2} by fixing $\Omega_H$ and changing scalar mass $\mu$. This figure shows that oscillation frequency $\omega_R$ of both states keep rising with $\mu$ whose larger value also enlarge the differences of $\omega_R$ between the two kind of states. Contrary to $\omega_R$, the $\omega_I$ of co-rotating states seems to monotonously decrease with $\mu$, while $\omega_I$ for counter-rotating states manifests peculiar behavior again, as we can observe some sharp $\omega_I$ increment followed by a larger decrement. These facts indicate that the co-rotating states with a larger mass will oscillate more rapidly and decay faster (to be specific, outside of the superradiant region), the same result for $\omega_R$ can also be concluded for counter-rotating states of which damping rate exhibit non-monotonic relationship with scalar mass. Further more, we compared the overtones of the two states in Table~\ref{tab4}. The higher overtone is related to higher oscillation frequency satisfying $\omega_R<\mu$ and larger $\omega_I$. However, for counter-rotating states, an exception is found at large overtone $n=14$ which has $\omega_R>\mu$ and smallest $\omega_I$ among the overtones in the Table.

At last we investigated the superradiant instability and its characteristics are reflected by Fig.~\ref{fig3} combined with Table~\ref{tab3}. When superradiant instability occurs, the black hole angular velocity $\Omega_H$, scalar mass $\mu$ and real component of QBS frequency $\omega_R$ must satisfy restrictions $\omega_R<m\Omega_H$ and $\omega_R<\mu<m\Omega_H+\Omega'$ where $\Omega'$ is some small positive value compared to $m\Omega_H$. Once $\omega_R>m\Omega_H$, the instability is absent since only negative $\omega_I$ is available. An interesting scenario is that when $\omega_R=m\Omega_H$, we can predict that $\omega_I=0$ which suggests that the states will neither grow nor decay, just as stationary scalar clouds formed around black hole. Under each angular velocity, the growth rate $\omega_I$ takes its maximum at $\mu\lesssim m\Omega_H$, and the maximum instability is not monotonously dependent on the angular velocity. When increasing $\Omega_H$ from $\Omega_H=0$ to $\Omega_H=1.22$, the maximum instability corresponding to each $\Omega_H$ grows with angular velocity, if we further increase $\Omega_H$, the maximum instability will start to decrease. This is a peculiar property of analog rotating black holes, unlike the case in  Kerr space time as it has been found that faster rotation creates greater instability in~\cite{Dolan:2007mj}. On the other hand, a larger winding number can significantly suppress the strength of instability. Thus, we can infer that a max instability $\omega_{Imax}$ related to a critical angular velocity exists in parameter space for $m=1$. In fact, the critical angular velocity is found to be about $\Omega_H\approx1.22$ and the corresponding $\omega_{Imax}\approx 1.13374\times 10^{-5}$.

It has been a decade since the first direct detection of GWs in human history, and since then we have  entered era of multi-messenger astronomy. During this ten years, great efforts have been put into the study related to the physics of GWs due to its promising applications, such as probing new physics beyond Standard Model by GWs from the ultralight bosons clouds produced by rapidly rotating black holes through superradiant instability. Therefore, superradiant instability of black holes plays an important role in new physics exploration. However, despite that the presence of superradiant instability is theoretically allowed, it has not been experimentally verified yet for astrophysical black holes, as it is a challenging task we are facing. Fortunately, analog black holes constructed in laboratory provide us an alternative accessible platform to theoretically and experimentally study black hole physics, including superradiant instability as what we have discussed in this work. Based on the theoretical analysis, the future experimental examination of superradiant instability of analog black holes  will unquestionably strengthen the prospect and confidence in observations of ultralight bosons clouds which help to facilitate the research of new physics. On the other hand, it should be noted that the superradiance has been observed by experiment in a photon superfluid~\cite{Braidotti:2021nhw}. Note that this experiment was performed for photon superfluid, although not for an analog black hole in photon-fluid, we believe that this achievement will facilitate the experimental test of superradiant instability of rotating black holes in photon-fluid model. 

\begin{acknowledgments}
This work is supported by National Natural Science Foundation of China under Grant No.12305071.
H.G. is also supported by the Institute for Basic Science (Grant~No.~IBS-R018-Y1).
\end{acknowledgments}

\bibliographystyle{JHEP}
\bibliography{References_Analog_BH1,References_Analog_BH2}

\end{document}